\documentclass[12pt,english,floatfix,nofootinbib,superscriptaddress,aps,prd,preprint]{revtex4}
\usepackage[utf8]{inputenc}
\usepackage{float}
\usepackage{placeins}
\usepackage{array}

\usepackage{tikz}
\usetikzlibrary{decorations.pathmorphing, arrows.meta}
\usepackage[dvips]{epsfig}

\usepackage{amsmath,amsthm,amsfonts,amssymb,amsbsy}
\usepackage{mathtools}
\usepackage{mathrsfs}
\usepackage{tensor}
\usepackage{esint}
\usepackage[makeroom]{cancel}
\usepackage{slashed}

\usepackage[english]{babel}
\usepackage{textcomp}
\usepackage{units}
\usepackage{xcolor}

\usepackage{hyperref}
\hypersetup{%
  colorlinks=true,
  breaklinks=true,
  citecolor=[rgb]{0,0.0,1.0},
  urlcolor=[rgb]{0.0,0.0,1.0},
  linkcolor=[rgb]{0,0.5,0.9}
}

\newcommand{\ie}{\begin{equation}}
\newcommand{\fe}{\end{equation}}
\newcommand{\se}{\begin{eqnarray}}
\newcommand{\ff}{\end{eqnarray}}

\begin{document}

\title{Carroll--Field--Jackiw term in nonlocal Lorentz-violating QED}

\author{J. R. Nascimento}
\email{jroberto@fisica.ufpb.br}
\affiliation{Departamento de Física, Universidade Federal da Paraíba, Caixa Postal 5008, 58051-970, João Pessoa, Paraíba,  Brazil} 

\author{A. Yu. Petrov}
\email{petrov@fisica.ufpb.br}
\affiliation{Departamento de Física, Universidade Federal da Paraíba, Caixa Postal 5008, 58051-970, João Pessoa, Paraíba,  Brazil} 

\author{P. J. Porfírio}
\email{pporfirio@fisica.ufpb.br}
\affiliation{Departamento de Física, Universidade Federal da Paraíba, Caixa Postal 5008, 58051-970, João Pessoa, Paraíba,  Brazil} 

\author{Ramires N. da Silva}
\email{rns2@academico.ufpb.br }
\affiliation{Departamento de Física, Universidade Federal da Paraíba, Caixa Postal 5008, 58051-970, João Pessoa, Paraíba,  Brazil} 

%%%%%%%%%%%%%%%%%%%%%%%%%%%%%%%%%%%%%%%%%%%%%%%%%%%%%%%%%%%%%%%%%%%%
%%%%%%%%%%%%%%%%%%%%%%%%%%%%%%%%%%%%%%%%%%%%%%%%%%%%%%%%%%%%
%%%%%%%%%%%%%%%%%%%%%%%%%%%%%%%%%%%%%%%%%%%%%%%%%%%%%%%%%%%%

\date{\today}

\begin{abstract}
We investigate the radiative generation of the Carroll--Field--Jackiw (CFJ) term in nonlocal Lorentz- and CPT-violating extensions of quantum electrodynamics. Nonlocality is introduced in the fermionic sector through entire form factors depending on the Dirac operator. Using the Fr\'echet and derivative expansions of the one-loop fermionic determinant, we derive the induced CFJ contribution and analyze its ultraviolet behavior after Wick rotation. For one class of form factors, the loop integrals are absolutely convergent at finite nonlocality scale $\Lambda$, without requiring an additional UV regulator. For another class of form factors, the CFJ coefficient is defined through an Abel regularization prescription. In both cases, the CFJ structure is preserved and its coefficient is continuously modulated by the ratio between the fermion mass and the nonlocality scale. Thus, nonlocality removes the routing-, surface-term-, and UV-evanescent ambiguities of the finite-$\Lambda$ loop amplitude, although UV convergence alone does not imply complete prescription independence of the finite functional determinant.
\end{abstract}

\maketitle

%%%%%%%%%%%%%%%%%%%%%%%%%%%%%%%%%%%%%%%%%%%%%%%%%%%%%%%%%%%%
%%%%%%%%%%%%%%%%%%%%%%%%%%%%%%%%%%%%%%%%%%%%%%%%%%%%%%%%%%%%

\section{INTRODUCTION}

Quantum electrodynamics (QED) is one of the most successful quantum field theories, providing extremely accurate predictions across a wide range of phenomena \cite{Peskin1995,Srednicki2007-li,Hann}. Nevertheless, several developments suggest that QED may be only an effective low-energy description of a more fundamental framework \cite{Green198,Polchinski1998,Gies}. In particular, possible departures from exact Lorentz/CPT invariance have attracted considerable attention, motivated by scenarios in quantum gravity, string theory, noncommutative geometry, and other beyond-Standard-Model settings \cite{AmelinoCamelia2005,Kostelecky2004,Mattingly2005}.

A widely studied class of QED extensions introduces Lorentz- and CPT-violating interactions in the fermionic sector \cite{Kostelecky:2001jc,Chung:1999gg,Andrianov:2001zj,Perez-Victoria:2001csb,Altschul:2004gs,Ferrari:2018tps}. A paradigmatic example is the axial-vector coupling $b_{\mu}\bar{\psi}\gamma^{\mu}\gamma_{5}\psi$, which modifies the fermion dynamics and may induce, through radiative corrections, a Carroll-Field-Jackiw (CFJ) term in the gauge sector, $ \mathcal{L}_{\rm CFJ}=k_{\mu}\epsilon^{\mu\nu\alpha\beta}A_{\nu}\partial_{\alpha}A_{\beta}$. The coefficient $k_{\mu}$ is proportional to the Lorentz-violating (LV) background vector $b_{\mu}$, i.e. $k_{\mu}=C\,b_{\mu}$, where $C$ is the CFJ coefficient. Since the pioneering works on LV QED (see, e.g., \cite{Carroll:1989vb,Jackiw:1999qq,Chung:1998jv,Chung:2001mb}), determining $C$ has been the subject of sustained interest. It is well known that the constant $C$ is sensitive to the regularization scheme, thereby leading to a long-standing debate about the origin and interpretation of this apparent ambiguity (see, e.g., \cite{Battistel2001,Bonneau2001,Altschul:2019eip}).

At the same time, nonlocal quantum field theories have emerged as promising frameworks for improving UV behavior while preserving key properties such as unitarity and gauge invariance. The paradigmatic paper addressing nonlocality in quantum field theory by Efimov \cite{Efimov:1967pjn} formulated key ideas of this concept. Further, nonlocal operators defined through entire functions have been investigated in various contexts, including modified gauge theories, quantum gravity, higher-derivative models and superfield theories (see, e.g., \cite{Tomboulis:1997gg,Biswas1,Biswas2,Modesto:2011kw,Modesto:2014lga,Modesto:2017sdr,Gama:2026ude} and references therein). Entire functions can suppress UV contributions without introducing additional poles in propagators, thereby avoiding ghost-like degrees of freedom.

Motivated by these developments, we study a nonlocal extension of Lorentz- and CPT-violating QED in which the nonlocal term is implemented by means of an entire function of the Dirac operator, rather than the usual d'Alembert operator \cite{Moffat:1990jj}. In fact, this choice is more general and convenient when we are dealing with spin-1/2 fermions because their kinetic term is proportional to the Dirac operator $\slashed{\partial}$. Such an idea was originally proposed in \cite{rami}. Earlier attempts to incorporate Lorentz symmetry breaking into nonlocal models can be found in \cite{Altschul:2004wr,Carone:2020zfk}. We follow the standard effective-field-theory strategy: the action is written as the sum of a Lorentz-invariant (here, nonlocal) Lagrangian and a small LV perturbation, in a close analogy with the LV Standard Model Extension (SME) \cite{Colladay,Colladay:1998fq}. Our aim is to compute the radiatively induced CFJ term in this setting and to assess how nonlocality modifies the effective action and the corresponding CFJ coefficient.

A central question is whether the nonlocal operator can naturally regulate the UV behavior of the loop integrals responsible for the induced CFJ term. As we already noted above, in conventional LV QED, the CFJ coefficient is typically sensitive to the chosen regularization prescription. We show that the situation changes significantly in the nonlocal framework considered here. For the first class of form factors, after Wick rotation the relevant loop integral becomes absolutely convergent due to exponential suppression. For the second class, the resulting oscillatory integral is defined through an Abel prescription.

Accordingly, the CFJ coefficient is finite in the first case directly from the convergent loop amplitude, whereas in the second case it is defined through the Abel prescription. In the present nonlocal setting, it is therefore essential to separate the UV convergence ensured by the entire form factor from the more general issue of how the functional determinant is normalized at finite order.

The paper is organized as follows. In Sec.~II, we introduce the nonlocal Lorentz- and CPT-violating extension of QED, discuss its basic properties, and derive the one-loop effective action. In Sec.~III, we evaluate the relevant loop integrals, perform the Wick rotation, and study the UV behavior for two representative choices of nonlocal form factors. We then extract the induced CFJ coefficients and compare them with earlier results in the literature. Finally, in Sec.~IV, we summarize our conclusions and comment on the role of nonlocality in Lorentz- and CPT-violating quantum field theories.

\section{Nonlocal LV QED}

In this section we define the nonlocal Lorentz- and CPT-violating extension of QED studied here. Nonlocality is introduced by multiplying the fermionic kinetic operator by an entire function of the gauge-covariant Dirac operator, $\slashed{D}=\gamma^{\mu}D_{\mu}$ with $D_{\mu}=\partial_{\mu}-i\mathbf{e}A_{\mu}$, while LV and CPT violation is encoded in a constant axial-vector background $b_{\mu}$. The corresponding Lagrangian is
\begin{equation}\label{lag}
\mathcal{L}=\bar{\psi}\big[i\slashed{D}\,f(\slashed{D})-m-\gamma_{5}\slashed{b}\big]\psi.
\end{equation}
Here $\psi$ is the Dirac field of mass $m$, $\mathbf{e}$ is the electric charge, and $A_{\mu}$ is the gauge field. This provides a Lorentz- and CPT-violating gauge-covariant generalization of the model considered in \cite{rami}. Choosing $f$ to be an entire function ensures analyticity and admits a power-series expansion,
\begin{equation}
f(\slashed{D})=\sum_{n=0}^{\infty} c_{n}\,\slashed{D}_{\!\Lambda}^{n},
\end{equation}
with $c_{0}=1$, to provide the correct local limit, and the dimensionless operator $\slashed{D}_{\!\Lambda}\equiv \slashed{D}/\Lambda$. The parameter $\Lambda$ sets the nonlocality scale: momenta $p\ll\Lambda$ probe the infrared (IR) regime where nonlocal effects are suppressed, while $p\gtrsim\Lambda$ probe the UV regime where the form factor significantly modifies loop integrals. The limit $\Lambda\to\infty$ corresponds to the local theory, $f\to 1$.

The quantum theory is defined by the generating functional
\begin{equation}\label{za}
Z[b,A]=\int\mathcal{D}\bar{\psi}\,\mathcal{D}\psi\,\exp\Big(i\int d^{4}x\,\mathcal{L}\Big).
\end{equation}

Substituting Eq. (\ref{lag}) into Eq. (\ref{za}), and integrating over the fermion fields, we obtain
\begin{equation}
Z[b,A]=\mathrm{Det}\big(i\slashed{D}f(\slashed{D})-m-\gamma_{5}\slashed{b}\big)=e^{i\Gamma[b,A]},
\end{equation}
which defines the one-loop effective action
\begin{equation}\label{seff}
\Gamma[b,A]=-i\,\mathrm{Tr}\,\ln\big[i\slashed{D}f(\slashed{D})-m-\gamma_{5}\slashed{b}\big].
\end{equation}
To compute the radiatively induced CFJ term, we must specify $f(\slashed{D})$; in the next section we analyze two representative exponential choices introduced in Ref.~\cite{rami}:
\begin{equation}\label{f1}
 f_{1}(\slashed{D})=e^{-\slashed{D}_{\!\Lambda}} 
\end{equation}
\begin{equation}\label{f2}
 f_{2}(\slashed{D})=e^{-i\slashed{D}_{\!\Lambda}}. 
\end{equation}

\subsection{Relation to derivative-expansion, proper-time, and thermal approaches}

Our strategy is closely aligned with earlier derivative-expansion analyses of the local CFJ problem, including \cite{Chung:1998jv}. The novelty of our study is not using the derivative expansion {\it per se}, but rather its application to the gauge-covariant nonlocal operator $i\slashed D f(\slashed D)$, whose Fr\'echet expansion produces momentum-dependent one-photon vertices as well as a genuine two-photon contact vertex. The proper-time formalism offers another convenient representation of the fermionic effective action \cite{Chung:2001mb}. In that approach, however, the exponential arises from a particular representation of the functional determinant, whereas the entire form factor is part of the microscopic dynamics itself.

There is also a structural analogy with finite-temperature LV QED \cite{Cervi:2001fg,Ebert:2004pq}, where the induced coefficient can be written as $C_T=C_{\rm ref}\, f\!\left(m/T\right)$, where $C_{\rm ref}$ is the CFJ reference coefficient. Likewise, the present nonlocal coefficient is controlled by the dimensionless ratio $m/\Lambda$, although the underlying physics differs. At finite temperature, the modification is a property of the thermal state, encoded by the Matsubara replacement and the preferred heat-bath four-velocity $u^\mu$. In the nonlocal theory, by contrast, $\Lambda$ enters the microscopic fermionic operator and modifies both propagators and interaction vertices. Thus, the analogy is limited to the presence of an additional physical scale that continuously modulates the induced parity-odd coefficient; the nonlocal deformation should not be interpreted as a thermal effect. It should also be noted that finite-temperature corrections are conventionally separated into a zero-temperature contribution and a thermal contribution that vanishes in the zero-temperature limit (cf. \cite{Cervi:2001fg,Ebert:2004pq}), whereas no analogous simple decomposition is available in the present nonlocal case.

\section{Specific Choices of the nonlocal Form Factor}

In this section, we perform the one-loop calculations for the nonlocal theories based on gauge-covariant generalizations of the two form factors introduced in Ref. \cite{rami}.

\subsection{Case 1: $f_{1}(\slashed{D})=e^{- \slashed{D}_{\!\Lambda}}$}\label{caso1}

Expanding Eq. (\ref{f1}) in powers of the gauge field and retaining terms up to second order (which suffices for the CFJ term) yields
\begin{equation}\label{expan}
 f_{1}(\slashed{D})=e^{-\slashed{\partial}_{\!\Lambda}}+\frac{\mathbf{e}}{\Lambda}e^{-\slashed{\partial}_{\!\Lambda}}\slashed{A}+\frac{\mathbf{e}^{2}}{2\Lambda^{2}}e^{-\slashed{\partial}_{\!\Lambda}}\slashed{A}\slashed{A},
\end{equation}
where $\slashed{\partial}_{\!\Lambda} \equiv \slashed{\partial}/\Lambda$. Because $\slashed{\partial}$ and $\slashed{A}$ do not commute, the naive Taylor expansion in Eq.~(\ref{expan}) is only schematic.

Although one could, in principle, employ the Baker--Campbell--Hausdorff (BCH) formula \cite{Hall2003,Blanes2009}, 
\begin{equation}\label{BHC}
e^X e^Y=e^{( X+Y+\frac12[X,Y]+\frac{1}{12}[X,[X,Y]]+\frac{1}{12}[Y,[Y,X]]+\cdots)},
\end{equation}
with $X=-\slashed{\partial}_{\!\Lambda}$ and $Y=i\mathbf{e}\slashed{A}/\Lambda$, this is not the most convenient framework for deriving the nonlocal Feynman vertices.

Indeed, if one truncates only in powers of the gauge field while retaining the full dependence on the nonlocality scale, the BCH expansion generates an infinite tower of nested commutators, $Y$, $[X,Y]$, $[X,[X,Y]]$, $[X,[X,[X,Y]]],\ldots$, all of which remain of the same order in the gauge field. The infinite hierarchy arises because each additional commutator introduces higher derivatives acting on the gauge field rather than additional powers of $A_\mu$.

If one simultaneously assumes a derivative expansion, $\partial_{\!\Lambda}\ll1$, then successive commutators become increasingly suppressed, $[X,Y]\sim \mathcal{O}\!\left( \partial{A}/\Lambda^2\right)$, $[X,[X,Y]] \sim\mathcal{O}\!\left( \partial^2 A/\Lambda^3 \right)\,\ldots$, and the BCH series may be truncated consistently. In this low-energy regime one finds, for example,
\begin{equation}
e^{X+Y}\simeq e^X \left( 1+Y+\frac{1}{2}Y^2-\frac{1}{2}[X,Y] \right),
\end{equation}
which already reproduces the linear and quadratic gauge-field vertices.

However, our goal is not to perform a low-energy expansion. Rather, we wish to preserve the complete nonlocal dependence on $\partial_{\!\Lambda}$ and expand only in powers of the external gauge field. In this situation the BCH representation Eq. (\ref{BHC}) becomes cumbersome, since the contribution at a fixed order in $A_{\mu}$ is distributed among infinitely many nested commutators.

The problem is therefore fundamentally different from the purpose for which the BCH formula is usually employed \cite{rami}, namely the rewriting of a product of exponentials of noncommuting operators as a single exponential containing an infinite commutator series \cite{Baker,Campbell189,Hausdorff1906}. Instead, we require a perturbative expansion of the single operator $e^{X+Y}$ around the unperturbed operator $X$, while preserving the exact ordering of the interaction operator $Y$.

For this purpose the Fr\'echet expansion, also known in this context as the Duhamel expansion, provides the natural framework \cite{Cartan2017-jv,Frechet}. The Fr\'echet derivative generates a perturbative expansion of $ e^{X+Y}$ directly in powers of $Y$, without reorganizing the result into nested commutators \cite{Higham2008,AlMohyHigham2009,HighamRelton2014}. Consequently, the ordered sequence of gauge-field insertions remains explicit at every order.

This feature is particularly advantageous for deriving Feynman rules. The first- and second-order terms immediately generate the linear and quadratic vertices, while preserving the exact operator ordering between the nonlocal exponential factors and the gauge-field insertions. In contrast, the BCH expansion reorganizes the same information into commutator structures, making the extraction of the ordered momentum-space vertices considerably less transparent.

The ordered representation is especially important in the calculation of the CFJ term, whose parity-odd contribution is sensitive to the ordering of Dirac matrices, nonlocal exponential operators, and external momenta. Therefore, although the BCH formula is mathematically equivalent and extremely useful in many operator algebraic applications, the Fr\'echet expansion provides a more direct, transparent, and operationally efficient route for constructing the Feynman rules of the present nonlocal theory.

Accordingly, we expand the nonlocal exponential operator in Eq.~(\ref{f1}) using the Fr\'echet expansion up to second order in the gauge field. This procedure generates a linear interaction vertex, $V_{1}$, and a quadratic (contact) vertex, $V_{2}$, which together account for all one-loop contributions relevant to the induced CFJ term. The Lagrangian then assumes the form
\begin{equation}
    \mathcal{L}= \bar{\psi}[i\slashed{\partial}e^{-\slashed{\partial}_{\!\Lambda}}-m+V_1+V_2+V_b]\psi,
\end{equation}
where the axial LV interaction $V_{b}=\!-\gamma_{5}\slashed{b}$ is treated perturbatively. Accordingly, it is not included in the free propagator; instead, we insert it as a single vertex and keep only terms linear in $b_{\mu}$. Following the  Fr\'echet expansion, the  linear vertex is given by
\begin{equation}\label{v1}
    V_1= \mathbf{e} \slashed{A} e^{-\slashed{\partial}_{\!\Lambda}}- \mathbf{e}\slashed{\partial}_{\!\Lambda} \int_{0}^{1} ds\, e^{-(1-s)\slashed{\partial}_{\!\Lambda}} \slashed{A}e^{-s\slashed{\partial}_{\!\Lambda}}.
\end{equation}

Similarly, the quadratic vertex, generated by the second-order term in the Fr\'echet expansion, reads
\begin{equation}\label{v2}
V_{2}= \frac{\mathbf{e}^2}{\Lambda}\slashed{A}\!\int_{0}^{1}\!\!ds \,e^{-(1-s)\slashed{\partial}_{\!\Lambda}}\slashed{A}e^{-s\slashed{\partial}_{\!\Lambda}}  -\frac{i \mathbf{e}^{2}}{\Lambda}\slashed{\partial}_{\!\Lambda}\!\int_{0}^{1} \!\!ds\!\int_{0}^{s}\!\!dt\,e^{-(1-s)\slashed{\partial}_{\!\Lambda}}\slashed{A}e^{-(s-t)\slashed{\partial}_{\!\Lambda}}\slashed{A}e^{-t\slashed{\partial}_{\!\Lambda}}.
\end{equation}
The auxiliary parameters $s$ and $t$ encode the operator ordering of gauge-field insertions inside the nonlocal exponential; they satisfy $0\le t\le s\le 1$ and disappear after the loop integrations are performed. 

The vertex (\ref{v2}) has no analogue in minimally coupled local QED, where the fermion--photon interaction contains only the usual three-point vertex, and is a direct consequence of the nonlocal form factor.

To expand the functional trace in Eq.~(\ref{seff}) in powers of the gauge field, we separate the operator into a free part and an interaction part. Defining $B_{0}=i\slashed{\partial}e^{-\slashed{\partial}_{\!\Lambda}}-m$, one may write
\begin{eqnarray}
\mathrm{Tr}\ln\big[i\slashed{D}f(\slashed{D})-m-\gamma_{5}\slashed{b}\big]
&=&\mathrm{Tr}\ln[B_{0}]+\mathrm{Tr}\ln[1+S_{0}(V_{1}+V_{2}+V_b)] \nonumber\\
&=&\mathrm{Tr}\ln[B_{0}]+\sum_{n=1}^{\infty} \frac{(-1)^{n+1}}{n} \mathrm{Tr}[(S_{0}(V_{1}+V_{2}+V_b))^{n}],
\label{trace}
\end{eqnarray}
where the free propagator is $S_{0}=B_{0}^{-1}$. The CFJ term appears at first order in $b_{\mu}$ and second order in the gauge field. At this order, the expansion in Eq.~(\ref{trace}) receives contributions from (i) the term with two insertions of the linear vertex and (ii) the contact term involving the quadratic vertex $V_2$. We denote these contributions by
\begin{equation}\label{gamaCFJ}
\Gamma_{{}_{\mathrm{CFJ}}}= i\,\mathrm{Tr}\big[S_{0}\gamma_{5}\slashed{b}\, S_{0}V_{1}S_{0}V_{1}\big] - i\,\mathrm{Tr}\big[S_{0}\gamma_{5}\slashed{b}\,S_{0}V_{2}\big].
\end{equation}

The simplification of the effective action in Eq.~(\ref{gamaCFJ}) relies on the cyclicity of the functional trace, which is valid when boundary terms vanish and the regularization preserves trace cyclicity \cite{Chung:1999gg, Andrianov:2001zj, Perez-Victoria:2001csb, Altschul:2004gs}. Moreover, at $\mathcal{O}(bA^2)$ different cyclic orderings in momentum space are related by a loop-momentum shift, so their difference can be written as a total derivative,
\begin{equation}
k_\mu\int_0^1d\xi \int\frac{d^4p}{(2\pi)^4} \frac{\partial}{\partial p^\mu} {\cal K}(p+\xi k,k;\Lambda).
\end{equation}
Since ${\cal K}(p)\sim e^{-3p/\Lambda}/p^4$, the corresponding surface term vanishes, and all cyclic orderings coincide. Therefore, at finite $\Lambda$ trace cyclicity is ensured by the convergence of the nonlocal kernel, rather than by imposing an independent momentum-routing prescription.

In particular, we do not assume any commutativity among $V_1$, $V_2$, and $V_b$, and the ordering of the nonlocal vertices is kept explicit throughout.

In momentum space, the quadratic (contact) vertex is obtained by summing over both possible orderings of the gauge field insertions,
\begin{equation}
V^{\mu\nu}_{2}=\mathbf{e}^{2}A_{\mu}(k_{1})A_{\nu}(k_{2})\Big[V^{\mu\nu}_{\mathrm{ord}}(p;k_{1},k_{2})+V^{\nu\mu}_{\mathrm{ord}}(p;k_{2},k_{1})\Big],
\end{equation}
where $k_{1}$ and $k_{2}$ are the external photon momenta and $p$ is the loop momentum. This form keeps the full nonlocal dependence on the external momenta and makes the Bose symmetry with respect to permutations of gauge fields  explicit,
\begin{equation}\label{simet}
V^{\mu\nu}_{2}(p;k_1,k_2) =V^{\nu\mu}_{2}(p;k_2,k_1).
\end{equation}

The ordered contribution is
\begin{equation}
\begin{aligned}\label{V2mu}
V^{\mu\nu}_{\mathrm{ord}}(p;k_{1},k_{2})&=\frac{\gamma^{\mu}}{\Lambda}\int_{0}^{1}ds\,e^{i(1-s)(\slashed{p}+\slashed{k}_{2})/\Lambda}\gamma^{\nu}e^{is\slashed{p}/\Lambda}
+\frac{1}{\Lambda^{2}}(\slashed{p}+\slashed{k}_{1}+\slashed{k}_{2})\\ &\quad \times \int_{0}^{1}ds\int_{0}^{s}dt\,e^{i(1-s)(\slashed{p}+\slashed{k}_{1}+\slashed{k}_{2})/\Lambda}\gamma^{\mu}e^{i(s-t)(\slashed{p}+\slashed{k}_{2})/\Lambda}\gamma^{\nu}e^{it\slashed{p}/\Lambda}.
\end{aligned}
\end{equation}

With this vertex, the relevant Dirac trace appearing in the second term of Eq.~(\ref{gamaCFJ}) can be written as
\begin{equation}\label{trace2}
\mathrm{Tr}\Big[S_{0}(p_{1})\gamma_{5}\slashed{b}\,S_{0}(p_{2})\big\{V^{\mu\nu}_{\mathrm{ord}}(k_{1},k_{2})+V^{\nu\mu}_{\mathrm{ord}}(k_{2},k_{1})\big\}\Big],
\end{equation}
where $p_{1}$ and $p_{2}$ denote the internal momenta immediately before and after the axial insertion, respectively. Since $b_{\mu}$ is taken to be constant, it carries no external momentum and hence does not alter the momentum running along the fermion line; therefore $p_{1}=p_{2}\equiv p$, and the external momentum conservation implies $k_{1}=-k_{2}\equiv k$.

We now introduce the shorthand notation
\begin{equation}
B\equiv\cos \rho,\qquad  M\equiv m  - i p\sin \rho\,,
\end{equation}
where we introduced the dimensionless momentum variable
\begin{equation}\label{rho}
 \rho \equiv p/\Lambda.
\end{equation}

The free fermion propagator for a line carrying momentum $p$ can be written compactly as
\begin{equation}\label{propag}
S_0=\frac{B\slashed{p} +M} {\Delta},
\end{equation}
where $\Delta=(Bp)^2 - M^{2}$, and $p\equiv \sqrt{p^2}$ denotes the Lorentz-invariant momentum magnitude. Substituting Eq.~(\ref{propag}) into Eq.~(\ref{trace2}), the parity-odd (pseudotensor) part of the trace can be written as
\begin{equation}
T_{V_2}^{\mu\nu}=\frac{1}{\Delta^{2}} \,\mathrm{Tr}\Big[ \left(B\slashed {p}+M\right) \gamma_5\slashed b \left(B\slashed {p}+M\right)V^{\mu\nu}_{2}\Big].
\end{equation}

Expanding the numerator yields
\begin{equation}\label{TV2}   
\begin{aligned}
T_{V_2}^{\mu\nu}=\frac{1}{\Delta^{2}} &\Big\{ M^{2}\,\mathrm{Tr}\!\left[ \gamma_5\slashed{b} \,V^{\mu\nu}_{2} \right] +MB\,\mathrm{Tr}\!\left[ \gamma_5\slashed b\,\slashed {p}\,V^{\mu\nu}_{2} \right] +BM\,\mathrm{Tr} \!\left[ \slashed {p}\gamma_5\slashed b\,V^{\mu\nu}_{2}\right] \\&+B^{2}\, \mathrm{Tr}\!\left[\slashed {p}\gamma_5\slashed b\,\slashed {p}\,V^{\mu\nu}_{2} \right]\Big\}.
\end{aligned}
\end{equation}

To isolate the pseudotensor contribution, we expand the quadratic vertex in the Clifford basis as
\begin{equation}
   V^{\mu\nu}_2 = V^{\mu\nu}+    V^{\mu\nu}_{\lambda}\gamma^{\lambda}+ \frac{1}{2} V^{\mu\nu}_{\lambda\alpha} \gamma^{\lambda}\gamma^{\alpha}+ \frac{1}{6} V^{\mu\nu}_{\lambda\alpha\beta} \gamma^{\lambda}\gamma^{\alpha}\gamma^{\beta} +\cdots
\end{equation}

The parity-odd contribution arises from terms whose Dirac traces reduce to structures containing $\gamma_5$ and four gamma matrices. The relevant trace reduces to
\begin{eqnarray}\label{tracedirac}
    \mathrm{Tr}[\gamma_5 \gamma^{\delta} \gamma_{\mu} \gamma_{\nu}\gamma_{\alpha}\gamma_{\delta} \gamma_{\beta}] = 2\,\mathrm{Tr}[\gamma_5 \gamma_{\mu}\gamma_{\nu}\gamma_{\alpha}\gamma_{\beta}]= 8i\epsilon_{\mu\nu\alpha\beta}, 
\end{eqnarray}  
where we used the four-dimensional identity $\gamma^{\delta}\gamma_{\mu}\gamma_{\delta}=-2\gamma_{\mu}$.

A potential subtlety is the continuation of the Dirac algebra to $D$ dimensions. For example,
\begin{equation}
\gamma_\rho\gamma^{\mu\nu\alpha}\gamma^\rho = (6-D)\gamma^{\mu\nu\alpha},
\end{equation}
so the parity-odd trace differs from its four-dimensional value by evanescent contributions of order $D-4$. Such terms can produce finite effects only if they are accompanied by UV poles. At finite $\Lambda$, however, the nonlocal integrals are smooth near $D=4$,
\begin{equation}
I_D(\Lambda)=I_4(\Lambda)+\mathcal{O}(D-4),
\end{equation}
and consequently
\begin{equation}
\lim_{D\rightarrow4}(D-4)I_D(\Lambda)=0.
\end{equation}
Thus no finite evanescent remnant survives, and the parity-odd Dirac trace may be evaluated directly in four dimensions for the convergent finite-$\Lambda$ amplitude.

Using Eq.~(\ref{tracedirac}), one can rewrite Eq.~(\ref{TV2}) as
\begin{equation}\label{tensor2}
T_{V_2}^{\mu\nu}= \frac{4i}{\Delta^2} \Big[\frac{1}{6}M^2\,b_\alpha V^{\mu\nu}_{\lambda\rho\beta} \epsilon^{\alpha\lambda\rho\beta} + MB\, b_\alpha p_{\beta} V^{\mu\nu}_{\lambda\rho} \epsilon^{\alpha\beta\lambda\rho} + B^2\, b_\alpha p_{\beta}p_{\rho} 
V^{\mu\nu}_{\lambda} \epsilon^{\alpha\beta\rho\lambda} \Big].
\end{equation}

The symmetry of Eq.~(\ref{simet}) ensures that Eq.~(\ref{tensor2}) is even under the change $(\mu,k_{1})\leftrightarrow(\nu,k_{2})$; hence the contact term has no antisymmetric component and cannot project onto the \emph{local} CFJ operator. Therefore, the induced CFJ coefficient comes entirely from the contribution with two linear vertices $V_{1}$.
\begin{figure}[H]
\centering
\begin{tikzpicture}[scale=0.7]
% External photon lines
\draw[decorate, decoration={snake}, thick] (0,2) -- (1.2,1);
\draw[decorate, decoration={snake}, thick] (4.8,1) -- (6,2);
% Fermion loop
\draw[thick, -{Latex}] (1.2,1) arc (180:0:1.8 and 1);
\draw[thick, -{Latex}] (4.8,1) arc (0:-180:1.8 and 1);
% Axial insertion
\draw[thick] (3,0) -- (3,0.35);
\filldraw[black] (3,0.35) circle (2pt);
\node[above] at (3,0.45) {$\gamma^5\slashed{b}$};
% External labels
\node at (-0.2,2.25) {$A_\mu(k)$};
\node at (6.2,2.2) {$A_\nu(-k)$};
\end{tikzpicture}
\vspace*{-0.6cm}
\caption{One-loop contribution to the induced CFJ term in the covariant nonlocal model. The photon vertices $A_\mu(k)$ and $A_\nu(-k)$ are nonlocal linear vertices $V_{1}$ generated by expanding $f_{1}(\slashed D)$. The bullet denotes an insertion of the zero-momentum axial operator $\gamma_5\slashed b$ on the fermion line.}
\label{fig:Feynman_diag}
\end{figure}
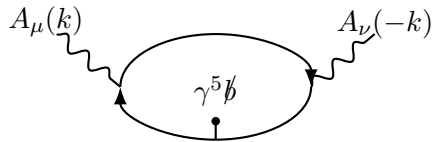

The derivative expansion framework is adopted following the method of Ref. \cite{Chung:1998jv} (see also \cite{Babu:1987rs,chan} for a comprehensive discussion of the method), based on the relation:
\begin{eqnarray}\label{expansion}
    \slashed{A}(x) &\rightarrow& \slashed{A} \Big( x-i\frac{\partial}{\partial p}  \Big)= \slashed{A}(x)-i \partial_{\mu} \slashed{A}(x) \frac{\partial}{\partial p_{\mu}}+ \cdots
\end{eqnarray}

Because $\partial_{\mu}(\gamma_{5}\slashed{b})=0$, the derivative expansion in Eq.~(\ref{expansion}) acts only on the momentum dependence of the free propagator. Keeping terms up to first order in derivatives and isolating the pieces relevant for the CFJ structure, one can write the trace in the first term of Eq.~(\ref{gamaCFJ}) in the form
\begin{equation}
\begin{aligned}
\mathrm{Tr}\Big[S_0  \gamma_5\slashed b\,S_0 \,\partial_\mu \!V_1 \frac{\partial S_0}{\partial p_\mu} V_1 +S_0\,\partial_\mu \!V_1 \frac{\partial S_0}{\partial p_\mu} \gamma_5\slashed b\,S_0V_1 +S_0\,\partial_\mu\! V_1 \frac{\partial S_0}{\partial p_\mu} V_1S_0\gamma_5\slashed b
\Big].
\end{aligned}
\label{gamafinal}
\end{equation}

Using Eq.~(\ref{tracedirac}) to evaluate the Dirac trace, one can write the parity-odd part of Eq.~(\ref{gamafinal}) in the compact form that yields the effective action contributing to the CFJ coefficient:
\begin{equation}\label{acaoefet}
  \Gamma_{{}_{\mathrm{CFJ}}}= -\frac{\mathbf{e}^2}{2}\int d^4x \int \frac{d^4 p}{(2\pi)^4} \frac{N}{\Delta^4},
\end{equation}
with the numerator $N$ given by
\begin{eqnarray}\label{N1}
\nonumber N\! &=& \!4i\Delta \Big\{ \epsilon^{\alpha\beta\mu\sigma} \left[3\cos\rho (m- i p \sin\rho)^2 + p^2\cos^3\!\rho  \right]  - 4  \cos^3\!\rho \,\epsilon^{\alpha\beta\mu\nu}p_{\nu}p^{\sigma} \\ &+& \!\!\frac{4i}{p} \sin\rho \cos\rho  (m- ip \sin\rho) \!\left[ \epsilon^{\alpha\beta\nu\sigma}p_{\nu}p^{\mu} - \epsilon^{\alpha\nu\mu\sigma}p_{\nu}p^{\beta}  +\epsilon^{\nu\beta\mu\sigma}p_{\nu}p^{\alpha}\right]   \! \!\Big\} b_{\sigma}\partial_{\mu} A_{\alpha} A_{\beta}
\end{eqnarray}
and the factor defining the denominator is
\begin{equation}\label{delta}
\Delta=(p\cos\rho)^2-\left(m-ip\sin \rho\right)^2,
\end{equation} 

Before performing the tensor reduction needed to extract the CFJ coefficient, note that the parity-odd amplitude involves tensor integrals of the form
\begin{equation}
\int d^4p\,p_\mu p_\nu \, H_i(p^2;\Lambda,m).
\end{equation}
After Wick rotation the kernels behave, for large momenta, as
\begin{equation}
H_i(p^2;\Lambda,m)\sim\frac{e^{-3p/\Lambda}}{p^6}, \qquad p\rightarrow\infty.
\end{equation}
Hence, for finite $0<\Lambda<\infty$, these integrals are absolutely convergent. Euclidean rotational invariance then implies
\begin{equation}
\int d^4p\,p_\mu p_\nu \, H_i(p^2)= \frac{\delta_{\mu\nu}}{4} \int d^4p\,p^2H_i(p^2),
\end{equation}
so the replacement $p_\mu p_\nu\rightarrow \frac{1}{4}\delta_{\mu\nu}p^2$ under the integral sign is an identity for a convergent integral, rather than an assumption of symmetric integration. 

Substituting Eqs.~(\ref{N1}) and (\ref{delta}) into Eq.~(\ref{acaoefet}), implementing the tensor reduction $p_{\mu}p_{\nu} \rightarrow \dfrac{p^{2}}{4} \eta_{\mu\nu}$, and isolating the coefficient of the CFJ structure, we obtain
\begin{equation}\label{cei}
C=\frac{i \mathbf{e}^2}{4\pi^2}\int_{0}^{\infty} d\rho\, \rho^3 \frac{cos\, \rho(\mu- i\rho\sin\,\rho)(3\mu-2i\rho\sin\,\rho)}{(\rho^2-\mu^2+2i\mu\rho\sin\,\rho)^3},
\end{equation}
with the dimensionless mass parameter
\begin{equation}\label{mu}
 \mu \equiv m/\Lambda.   
\end{equation}

After Wick rotating Eq.~(\ref{cei}), we can write the result in Euclidean space as
\begin{equation}\label{keucli}
C=C^{\mathrm{loc}}\,f(\mu),
\end{equation}
where
\begin{equation}\label{before}
C^{\mathrm{loc}}=\frac{3 \mathbf{e}^2}{16\pi^2}
\end{equation}
is the local result of Ref.~\cite{Jackiw:1999qq}, and
\begin{equation}\label{funcaodeform}
f(\mu)=\frac{4}{3}\int_{0}^{\infty} d\rho\, \rho^3 \, \frac{\cosh\rho\,(\mu+\rho\sinh\rho)\,(3\mu+2\rho\sinh\rho)}{\big[(\rho\cosh\rho)^2+(\mu+\rho\sinh\rho)^2\big]^3}
\end{equation}
encodes the deviation from the local coefficient in Eq.~(\ref{before}) through a dimensionless deformation function.

At this stage we must check the momentum-space surface term,
\begin{equation}
{\cal S}_{\mu\nu} = \int\frac{d^4p}{(2\pi)^4} \, \frac{\partial}{\partial p^\mu} \left[p_\nu F(p^2)\right] = \frac{\delta_{\mu\nu}}{32\pi^2} \lim_{R\rightarrow\infty}R^4F(R^2).
\end{equation}
For the nonlocal kernel we have
\begin{equation}
F(R^2)\sim\frac{e^{-3R/\Lambda}}{R^5}\,\,\, \Longrightarrow\,\, {\cal S}_{\mu\nu}=0,
\end{equation}
so the surface term vanishes.

The same exponential damping also guarantees invariance under loop-momentum shifts. Indeed,
\begin{equation}
I(a)-I(0) = a_\mu\int_0^1 d\xi\int d^4p\, \frac{\partial}{\partial p_\mu} \, {\cal I}(p+\xi a),
\end{equation}
and the associated boundary term is zero because ${\cal I}(p)\sim e^{-3p/\Lambda}/p^4$. Hence, for finite $\Lambda$, the amplitude is independent of the momentum routing.

This should be contrasted with the local theory. Setting $f_{1}=1$ before integration yields
\begin{equation}
\Upsilon_{\mu\nu} = \int\frac{d^4p}{(2\pi)^4} \, \frac{\partial}{\partial p^\mu} \left[ \frac{p_\nu}{(p^2+m^2)^2} \right].
\end{equation}
With a four-dimensional spherical cutoff one obtains
\begin{equation}\label{surfaceterm}
\Upsilon_{\mu\nu}^{\rm sym} = \frac{\delta_{\mu\nu}}{32\pi^2},
\end{equation}
whereas a shift-invariant prescription can consistently set $\Upsilon_{\mu\nu}=0$. The local surface term, and therefore the finite CFJ contribution, is prescription dependent. For this reason, the local value obtained with the prescription adopted in our calculation, Eq.~(\ref{before}), should not be interpreted as a universal regulator-independent coefficient.

We thus conclude that, at finite $\Lambda$, the form factor removes the ambiguities associated with symmetric tensor reduction, momentum routing, surface terms, and UV-enhanced evanescent contributions in the loop amplitude. This, however, does not by itself guarantee uniqueness of the functional determinant defining the one-loop effective action in Eq.~(\ref{seff}). Distinct admissible definitions of this infinite-dimensional determinant may still differ by finite local terms, including
\begin{equation}
\Delta\Gamma_{{}_{\rm CFJ}} = \Delta C \int d^4x\, b_\mu\epsilon^{\mu\nu\rho\sigma} A_\nu\partial_\rho A_\sigma.
\end{equation}
Thus, while nonlocality eliminates routing-, surface-term-, symmetric-integration-, and UV-evanescent ambiguities in the finite-$\Lambda$ loop computation, UV convergence alone does not establish full prescription independence of the finite normalization of the CFJ operator.

To avoid ambiguities in the analytic continuation, we Wick rotate Eq.~(\ref{cei}) (metric signature $(+---)$) at the level of the Dirac operator, before rationalizing the propagator. In Euclidean space the denominator in Eq.~(\ref{funcaodeform}) is manifestly positive definite, so the propagator has no singularities over the integration domain, ensuring an unambiguous evaluation of the finite-$\Lambda$ loop integral entering Eq.~(\ref{keucli}) within the adopted definition of the effective action.

The Euclidean coefficient in Eq.~(\ref{keucli}) is real and finite. Convergence follows from the asymptotic behavior of the integrand in Eq.~(\ref{funcaodeform}) in the IR and UV regimes,
\begin{equation}\label{funcaoiruv}
f(\mu)_{{}_\mathrm{IR}} \sim \int_{0}^{\epsilon} d\rho\,\frac{3\rho^3}{\mu^4} \qquad \text{and} \qquad f(\mu)_{{}_\mathrm{UV}}\sim \int_{\sigma}^{\infty} d\rho\, \dfrac{4e^{-3\rho}}{\rho},
\end{equation}
where $\epsilon$ and $\sigma$ delimit the IR and UV domains, respectively.

In generic nonlocal theories, analytic continuation to Euclidean space can be subtle, since form factors may grow exponentially in certain directions of the complex momentum plane \cite{Efimov,Talaganis,Buoninfante}. In the present case, the choice in Eq.~(\ref{f1}) provides sufficient damping. This can be verified a posteriori from the large-momentum behavior of the Euclidean integrand in Eq.(\ref{funcaoiruv}): its exponential decay implies that the Minkowski-to-Euclidean contour deformation receives no contribution from infinity. The Euclidean formulation is therefore justified for evaluating the induced CFJ term.

Taking the local limit $\Lambda\rightarrow\infty$ (equivalently $\mu\to 0$) \emph{before} performing the loop integration removes the nonlocal deformation and reduces the theory to local LV QED, reproducing the usual (regularization-dependent) local result in Eq.~(\ref{before}). If, instead, one performs the nonlocal loop integral first and only then takes the local limit, the result is given by Eq.~(\ref{keucli}). This shows explicitly that the local limit and loop integration need not commute:
\begin{equation}\label{ncomuta}
\lim_{\mu\to 0} \int d\rho\,K\left(\rho,\mu\right) \ne \int d\rho\, \lim_{\mu\to 0}K\left(\rho,\mu\right),
\end{equation}
where $K(\rho,\mu)$ denotes the kernel of Eq.~(\ref{funcaodeform}).

While Eq.~(\ref{ncomuta}) resembles the well-known scheme dependence of the CFJ coefficient discussed in \cite{Jackiw:1999qq} and related works, the origin here is different: the nonlocal UV structure effectively provides an intrinsic UV damping mechanism. Consequently, the finite result retains a controlled dependence on how the local limit is implemented, offering a nonlocal analogue of regularization-scheme dependence in local LV QED. 

It is important to emphasize that the dimensionless mass parameter $\mu$ encodes the ratio of two physical scales in the model (see Eq.~(\ref{mu})) and should not be confused with the loop-momentum variable $\rho$ of Eq.~(\ref{rho}). Moreover, taking the local limit $\Lambda\to\infty$ (equivalently $\mu\to0$) in Eq.~(\ref{funcaodeform}) does \emph{not} reduce the integration range: one still integrates over $0\le\rho<\infty$. Since the physical Euclidean momentum is $p=\Lambda\rho$, fixed $\rho$ probes arbitrarily large $p$ as $\Lambda$ increases; hence $\mu\to0$ does not isolate an IR sector, and the integral continues to receive contributions from momenta $p\sim\Lambda$, accounting for the noncommutativity in Eq.~(\ref{ncomuta}).

In short, the mismatch between the two procedures is a UV effect: the momentum region $p\sim\Lambda$ contributes a finite amount when the nonlocal integral is performed first, but it is absent if one replaces the nonlocal operator by its local approximation from the outset. This highlights how entire-function nonlocality can simultaneously regulate UV behavior and leave a finite imprint on radiatively induced Lorentz- and CPT-violating terms.

\begin{figure}[H]
    \centering
    \includegraphics[width=0.5\textwidth]{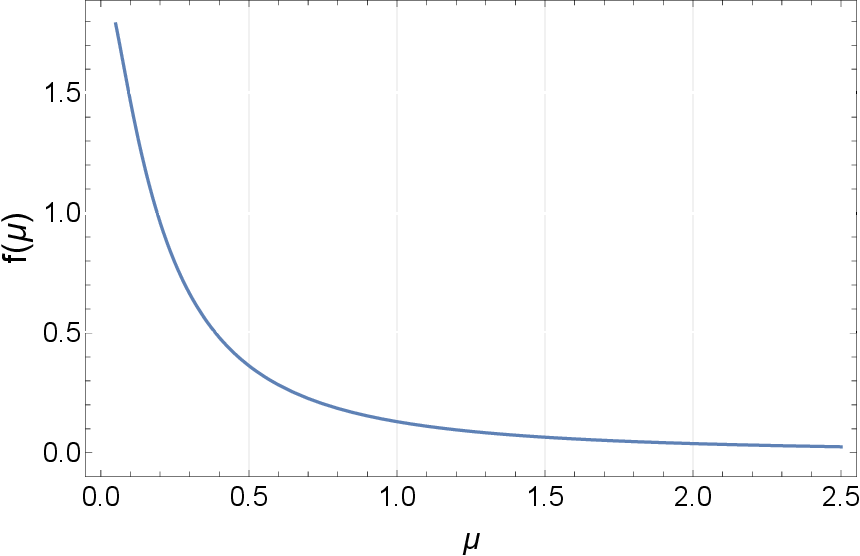}
    \vspace*{-0.6cm}
        \caption{Dimensionless nonlocal deformation function $f(\mu)$.}
    \label{fig:deforma}
\end{figure}

Fig.~\ref{fig:deforma} shows the deformation function given by Eq. (\ref{funcaodeform}), which quantifies how nonlocality changes the induced CFJ coefficient relative to the local value given by Eq. (\ref{before}). As $\mu$ increases, $f(\mu)$ decreases monotonically, showing that the radiative generation of the CFJ term is progressively suppressed when the fermion mass becomes comparable to (or larger than) the nonlocality scale. The crossing $f(\mu)=1$ occurs at $\mu\simeq 0.127$, where the nonlocal result matches the local one, cf. Eq.~(\ref{keucli}); for $\mu>0.127$ the induced coefficient is smaller than the local value.

For $\mu > 2$, $f(\mu)$ tends rapidly to zero, reflecting the strong suppression of the loop contribution when the fermion mass exceeds the nonlocality scale. Thus, the nonlocal operator does not introduce new gauge structures; it modifies only the magnitude of the CFJ coefficient through the ratio $m/\Lambda$.
\begin{figure}[H] %\vspace*{-2cm}
    \centering
    \includegraphics[width=0.5\textwidth]{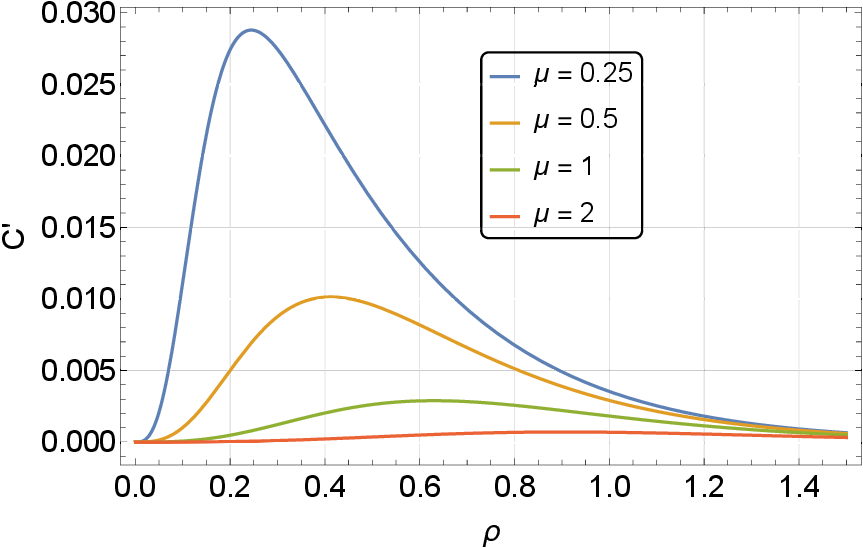}
    \vspace*{-0.6cm}
        \caption{Euclidean integrand contributing to $C^{\prime}=C/\mathbf{e}^2$ as a function of the dimensionless loop momentum $\rho$, for different $\mu$ values: $\mu=2$ (red line), $\mu=1$ (green line), $\mu=0.5$ (yellow line) and $\mu=0.25$ (blue line).}
    \label{fig:coeffi}
\end{figure}

Fig.~\ref{fig:coeffi} shows the (Euclidean) integrand that contributes to $C'=C/\mathbf{e}^{2}$ for several values of $\mu$. In all cases the integrand vanishes at $\rho=0$, rises quickly to a single maximum at low $\rho$, and then falls off exponentially for large $\rho$.

The peak height decreases as $\mu$ increases. Physically, as the fermion mass becomes larger relative to the nonlocality scale, the form factor suppresses the loop contribution responsible for the LV CFJ term.

For all plotted values of \(\mu\), the integrand is already negligible for $\rho\gtrsim 1.4$. Hence the dominant contribution to $C$ comes from a finite momentum window $\rho\lesssim 1.4$, and the nonlocal operator effectively acts as a UV regulator after Wick rotation.

Overall, nonlocality leaves the CFJ tensor structure unchanged and affects only its magnitude through the ratio $\mu=m/\Lambda$, with larger $\mu$ leading to stronger suppression.

%%%%%%%%%%%%%%%%%%%%%%%%%%%%%%%%%%
%%%%%%%%%%%%%%%%%%%%%%%%%%%%%%%%%%
\subsection{Case 2: $f_{2}(\slashed{D})=e^{-i\slashed{D}_{\!\Lambda}}$}

Following the same steps as in Case 1, one can write the Lagrangian for the oscillatory form factor $f_{2}$ as
\begin{equation}
    \mathcal{L}= \bar{\psi}[i\slashed {\partial}e^{-i\slashed{\partial}_{\!\Lambda}} - m + V_1+V_2+V_b]\psi.
\end{equation}

The one-loop effective action in Eq.~(\ref{acaoefet}) and the denominator factor $\Delta$ in Eq.~(\ref{delta}) retain the same structure as in the first model. The dependence on the nonlocal form factor enters through the vertices $V_{1}$ and $V_{2}$ and is therefore encoded in the numerator, which now reads
\begin{eqnarray}\label{N2}
\nonumber   N&=& \! 4i\Delta \Big\{\epsilon^{\alpha\beta\mu\sigma} \left[3\cosh\rho(m- i p \sinh\rho)^2 + p^2\cosh^3\!\rho \right]  - 4  \cosh^3\!\rho\,\epsilon^{\alpha\beta\mu\nu}p_{\nu}p^{\sigma} \\ &+& \!\!\!\frac{4i}{p} \sinh\rho \cosh\rho  (m \!-\! ip \sinh\rho)\! \!\left[   \epsilon^{\alpha\beta\nu\sigma}p_{\nu}p^{\mu} \!- \!\epsilon^{\alpha\nu\mu\sigma}p_{\nu}p^{\beta}  \!+\!\epsilon^{\nu\beta\mu\sigma}p_{\nu}p^{\alpha}\right] \!\!\Big\}   b_{\sigma}\partial_{\mu} A_{\alpha} A_{\beta}.
\end{eqnarray}

The corresponding nonlocal vertices are
\begin{eqnarray}
V_1&=& \mathbf{e}\,\slashed A\,e^{-i\slashed\partial_{\!\Lambda}} - i \mathbf{e} \slashed\partial_{\!\Lambda} \int_0^1 ds\, e^{-i(1-s)\slashed\partial_{\!\Lambda}} \slashed A e^{-is\slashed\partial_{\!\Lambda}},\\
\label{V1b}
\nonumber V_2&=& -\frac{\mathbf{e}^2}{\Lambda} \slashed A \int_0^1 ds\,e^{-i(1-s)\slashed\partial_{\!\Lambda}} \slashed A e^{-is\slashed\partial_{\!\Lambda}} \\ 
&+&\frac{i \mathbf{e}^2}{\Lambda} \slashed\partial_{\!\Lambda} \int_0^1 ds \int_0^s dt\,e^{-i(1-s)\slashed\partial_{\!\Lambda}}
\slashed A e^{-i(s-t)\slashed\partial_{\!\Lambda}} \slashed A e^{-it\slashed\partial_{\!\Lambda}}.
\label{V2b}
\end{eqnarray}

As in Case 1, the Bose-symmetrized contact vertex $V_{2}$ does not project onto the local CFJ operator, so the CFJ coefficient is determined by the diagram with two $V_{1}$ insertions. The analytic continuation of Eq.~(\ref{N2}) leads to an oscillatory UV behavior, which makes the Wick rotation more delicate and requires a prescription to define the loop integral. Applying Wick rotation yields
\begin{equation}\label{Ceucli2}
C=\frac{\mathbf{e}^2}{4\pi^2}\int_{0}^{\infty} d\rho\, \rho^3 \frac{cos\, \rho(\mu+\rho\sin\,\rho)(3\mu+2\rho\sin\,\rho)}{(\rho^2+\mu^2+2\mu\rho\sin\,\rho)^3}
\end{equation}  

The integral in Eq.~(\ref{Ceucli2}) is oscillatory in the UV and is not absolutely convergent. 
We therefore define the coefficient through the Abel regularization prescription \cite{Hardy1949,Copson1965,Saharian2024}. We introduce an exponential damping factor $e^{-\eta\rho}$, evaluate the integral for $\eta>0$, and finally take $\eta\to0^{+}$:
\begin{equation}
C=\frac{\mathbf{e}^2}{4\pi^2} \lim_{\eta\rightarrow 0^{+}}\int_{0}^{\infty} d\rho\, e^{-\eta\rho}\,K(\rho,\mu),
\end{equation}
where $K$ denotes the kernel of Eq.~(\ref{Ceucli2}). This procedure provides a controlled Abel definition of the oscillatory integral.

If the local limit $\Lambda\rightarrow\infty$ is taken \emph{before} loop integration, then $f_{2}\to 1$ and the theory reduces to local LV QED. In that procedure one recovers the local value adopted in the present prescription \cite{Jackiw:1999qq},
\begin{equation}\label{local2}
    C^{\mathrm{(loc)}}=\frac{3\mathbf{e}^{2}}{16\pi^{2}},
\end{equation}
without any need for Abel regularization.

If instead one performs the Abel-regularized loop integration first, the regularized deformation function is defined as
\begin{equation}
f_{\eta}(\mu) = \frac{4}{3} \int_{0}^{\infty} d\rho\, e^{-\eta\rho}\rho^{3} \frac{\cos\rho\,(\mu+\rho\sin\rho) (3\mu+2\rho\sin\rho) }{ \left(\rho^2+\mu^2+2\mu\rho\sin\rho \right)^3 }, \qquad \eta>0.
\end{equation}
The Abel-defined deformation function is then
\begin{equation}
f(\mu)=\lim_{\eta\to0^{+}}f_{\eta}(\mu),
\end{equation}
so that the CFJ coefficient can be written as
\begin{equation}\label{after2}
C=C^{(\mathrm{loc})}f(\mu).
\end{equation}

\begin{figure}[H]
\centering
\includegraphics[width=0.5\textwidth]{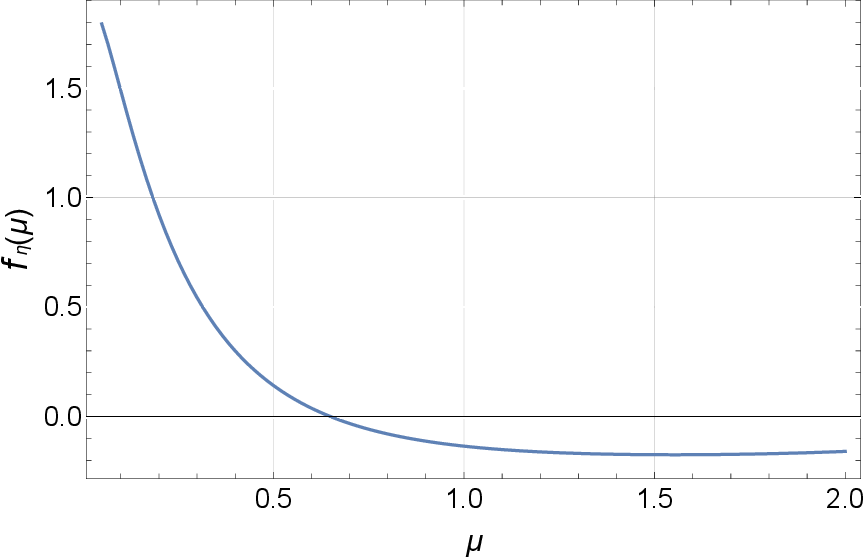} \vspace*{-0.6cm}
\caption{Dimensionless Abel-regularized deformation function $f_{\eta}(\mu)$ for $\eta=0.03$.}
\label{fig:deform2}
\end{figure}

Fig.~\ref{fig:deform2} displays $f_{\eta}(\mu)$ for a representative Abel parameter $\eta=0.03$. For $\mu\ll 1$ ($\Lambda\gg m$), 
the dependence of $f_{\eta}$ on the ratio $m/\Lambda$ becomes weak; however, because the Abel-regularized integration and the local limit do not generally commute, $f_{\eta}(\mu)$ need not approach unity for fixed finite $\eta$; only after taking the Abel limit $\eta\to0^+$ does one obtain the Abel-defined deformation function $f(\mu)$. As $\mu$ increases, $f_\eta(\mu)$ decreases and eventually becomes negative. This sign change is a direct consequence of the oscillatory Euclidean integrand: as $\mu$ grows, cancellations between positive and negative momentum regions become stronger and may flip the overall sign of the induced coefficient. This behavior reflects the oscillatory nonlocal dynamics rather than an inconsistency.

Since the $\mu\to0$ limit of Eq.~(\ref{after2}), taken after the Abel-defined loop integration, does not coincide with Eq.~(\ref{local2}), the local limit and loop integration do not commute: the loop integral retains a finite contribution from momentum regions in which the nonlocal operator modifies the UV behavior, even if the local limit is imposed afterward.

\begin{figure}[!htpb]
\centering
\includegraphics[width=0.5\textwidth]{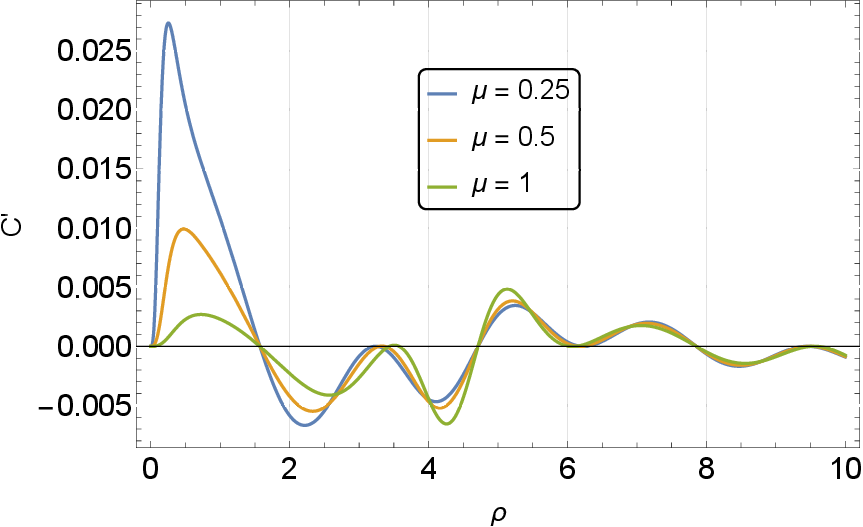} \vspace*{-0.6cm}
\caption{Abel-regularized (for $\eta=0.03$) Euclidean integrand contributing to $C^{\prime}=C/\mathbf{e}^2$ as a function of the dimensionless loop momentum $\rho$, for different $\mu$ values: $\mu=1$ (green line), $\mu=0.5$ (yellow line) and $\mu=0.25$ (blue line).}
\label{fig:abel}
\end{figure}

Fig.~\ref{fig:abel} shows the Abel-regularized ($\eta=0.03$) Euclidean integrand contribution to $C^{\prime}=C/\mathbf{e}^2$ as a function of $\rho$ for three values of $\mu$. The integrand exhibits damped oscillations whose amplitude decreases with increasing $\rho$. This is qualitatively different from Case 1: here the oscillatory contribution is defined through a controlled Abel summation procedure rather than by absolute UV convergence.

%%%%%%%%%%%%%%%%%%%%%%%%%%%%%%%%%%%%%%%%%%%%%%%%%%%%%%%%%%%%
%%%%%%%%%%%%%%%%%%%%%%%%%%%%%%%%%%%%%%%%%%%%%%%%%%%%%%%%%%%%

\FloatBarrier
\section{Conclusions}

We studied a nonlocal Lorentz- and CPT-violating extension of QED in which the fermionic kinetic operator is dressed by an entire function of the gauge-covariant Dirac operator. Our focus was the one-loop generation of the CFJ term and, in particular, how the nonlocal form factor reshapes the UV behavior of the effective action. Although the present calculation is performed within a genuinely nonlocal extension of QED, it shares important technical features with earlier approaches to the CFJ term and can be compared with other calculational schemes, such as the derivative expansion of Ref. \cite{Chung:1998jv} and the proper-time method of Ref. \cite{Chung:2001mb}.

For the two representative exponential form factors analyzed here, the induced parity-odd contribution remains gauge invariant and retains its usual tensor structure, but its coefficient is rendered finite through two qualitatively different mechanisms. In the first model, the Wick-rotated loop integral is exponentially suppressed at large Euclidean momenta, so the nonlocal operator provides an intrinsic UV damping mechanism. The result is a finite deformation of the local CFJ coefficient, free from momentum-routing, surface-term, symmetric-integration, and UV-evanescent ambiguities at finite $\Lambda$. There is a limited structural analogy with the finite-temperature case, in the sense that an additional physical scale continuously modulates the induced CFJ coefficient, see e.g. \cite{Cervi:2001fg,Ebert:2004pq} for finite-temperature results.

In the second model, the UV behavior is oscillatory and the loop integral is not absolutely convergent. Nevertheless, an Abel prescription provides a well-defined finite CFJ coefficient by damping the oscillations and defining the corresponding Abel limit. Thus, while both nonlocal theories lead to finite radiative corrections, they do so either through genuine absolute UV convergence (Case 1) or through an Abel-defined oscillatory integral (Case 2).

Overall, our results show that nonlocality offers a systematic way to tame UV sensitivity in radiatively induced topological terms without modifying their gauge structure. This strengthens the case for using entire-function nonlocal operators as a natural mechanism for improving and controlling the UV behavior in Lorentz- and CPT-violating extensions of gauge theories and motivates further applications to other LV radiative effects.

A natural continuation of this study would be to develop nonlocal generalizations of other LV theories, in particular LV gravity. We plan to perform these studies in forthcoming papers.

\section*{Acknowledgments}
The authors thank the Conselho Nacional de Desenvolvimento Cient\'ifico e Tecnol\'ogico (CNPq). Albert Yu. Petrov thanks CNPq (grant No. 303777/2023-0). Paulo J. Porf\'irio thanks CNPq (grant No. 307628/2022-1).

%%%%%%%%%%%%%%%%%%%%%%%%%%%%%%%%%%%%%%%%%%%%%%%%%%%%%%%%%%%%
%%%%%%%%%%%%%%%%%%%%%%%%%%%%%%%%%%%%%%%%%%%%%%%%%%%%%%%%%%%%

% --- BibTeX option ---
% A BibTeX database file has been added as references.bib.
% If you want to use BibTeX instead of the manual thebibliography below, replace
% everything from \begin{thebibliography} to \end{thebibliography} with:
%   \bibliographystyle{apsrev4-1}
%   \bibliography{references}
% --- end BibTeX option ---

\end{document}